\documentclass[aps,prb,preprint,footinbib]{revtex4-1}

\usepackage{amsmath}  
\usepackage{amsfonts} 
\usepackage{graphicx} 
\usepackage{multirow}

\newcommand{\figurewidth}{0.9\textwidth}
\begin{document}


\title{Wiedemann-Franz law demonstration in a student practicum}

\author{V.N.Glazkov}
\email{glazkov@kapitza.ras.ru} 

\affiliation{Moscow Institute of Physics and Technology, 141700
Dolgoprudny, Russia}

\affiliation{P.L.Kapitza Institute for Physical Problems RAS, Kosygin
str. 2, 119334 Moscow, Russia}

\author{L.Ginzburg}
\affiliation{Moscow Institute of Physics and Technology, 141700
Dolgoprudny, Russia}

\author{A.Orlov}
\affiliation{Moscow Institute of Physics and Technology, 141700
Dolgoprudny, Russia}


\date{\today}

\begin{abstract}
Wiedemann-Franz law is a prediction of electronic theory of electric and thermal conductivity in metals, which states that a Lorenz ratio $L=\kappa/(\sigma T)$, where $\kappa$ is a thermal conductivity, $\sigma$ --- electric conductivity and $T$ --- absolute temperature, is a universal constant in certain cases. We present here a simple experimental setup to verify this prediction in a teaching experiment.
\end{abstract}

\maketitle 

\section{Introduction} 

Simple electronic theory of electrical conductivity (Drude model) explains electrical conductivity $\sigma$ of  electron gas in a metal as its drift in an applied electric field. Mobility of the charge carriers in this process is determined by a certain relaxation time $\tau_\mathrm{e}$, which is a  mean free time before electron scattering processes limiting the electric conductivity. This model finally results in a well-known Drude formula for electric conductivity \cite{kittel}:

\begin{equation}\label{eqn:sigma-drude}
\sigma=\frac{n e^2 \tau_\mathrm{e}}{m}
\end{equation}
\noindent here $n$ is a charge carriers concentration, $e$ is a carrier charge (i.e. electron charge), $\tau_e$ is a relaxation time   and $m$ is an effective mass of the carrier.

Thermal conductivity $\kappa$ of the degenerate electron gas can be found in a gas approximation as $\kappa=\frac{1}{3} v C^{(V)} l_\mathrm{th}=\frac{1}{3} v^2 C^{(V)} \tau_\mathrm{th}$, where $v$ is a characteristic velocity of gas particles (Fermi velocity here), $C^{(V)}$ is a specific heat per unit volume and $l_\mathrm{th}$ and $\tau_{\mathrm{th}}$ are a mean free path and a relaxation time for the scattering processes limiting the thermal conductivity. Combining it with the known result  for electronic gas specific heat \cite{kittel}
$C^{(V)}=\frac{\pi^2}{2}  n k_B \frac{k_B T}{E_F}$, where $E_F$ is a Fermi energy, $T$ is the absolute temperature (in Kelvins) and $k_\mathrm{B}$ is the Boltzman constant, one obtains:
\begin{equation}\label{eqn:kappa-gas}
\kappa=\frac{\pi^2}{3} \frac{n \tau_\mathrm{th}}{m} k_\mathrm{B}^2 T.
\end{equation}

These results of free electron model can be combined in such a way that all material dependent parameters will cancel. Namely, if relaxation times $\tau_\mathrm{e}$ and $\tau_\mathrm{th}$ coincide, then the ratio $\kappa/(\sigma T)$ is a universal constant called Lorenz number

\begin{equation}\label{eqn:Lorenz}
L=\frac{\kappa}{\sigma T}=\frac{\pi^2}{3} \left(\frac{k_\mathrm{B}}{e} \right)^2\approx 2.44 \times 10^{-8} \frac{\mathrm{W}\cdot\mathrm{Ohm}}{\mathrm{K}^2}
\end{equation}

This result is called Wiedemann-Franz law.\footnote{It is interesting to note that G.Wiedemann and R.Franz has established the Wiedemann-Franz law empirically well before the electronic theory of metals was developed and even well before the discovery of electron by J.J.Tomson! In their paper from 1853 \protect{\cite{wf}} they noted that at the same temperature thermal conductivity and electrical conductivity of metals scales. This observation lead them to the conclusion that thermal and electric transport is metals are related. Proportionality of $\kappa/\sigma$ ratio to the absolute temperature was later noted (also empirically) by L.Lorenz in 1872.\protect{\cite{lorenz}}} Its validity depends on the validity of $\tau_\mathrm{e}=\tau_\mathrm{th}$ condition, which in turn depends on the electron scattering processes relevant for thermal and electric conductivity processes.

In fact, Wiedemann-Franz law is more general than the free electron model mentioned above. It can be shown, see e.g. Ref.\onlinecite{abrikosov}, that for the isotropic Fermi surface thermal conductivity $\kappa=\frac{\pi^2}{9} T \left(V^2 \tau_\mathrm{th} D(E)\right)_{E=E_\mathrm{F}}$ and electric conductivity $\sigma=\frac{1}{3}e^2 \left(V^2 \tau_\mathrm{e} D(E)\right)_{E=E_\mathrm{F}}$, here $\tau_\mathrm{e}$ and $\tau_\mathrm{th}$ are the relaxation times for the  charge and thermal  transport as defined within $\tau$-approximation of kinetic equation, $V$ is the electron group velocity and $D(E)$ is the density of states, index $(...)_{E=E_\mathrm{F}}$ means that corresponding quantity is calculated on the Fermi surface. Again, if relaxation times are the same then the ratio $\kappa/(\sigma T)$ is a universal constant.

Condition $\tau_\mathrm{e} = \tau_\mathrm{th}$ does not hold in general: it requires the scattering processes limiting thermal and charge transport to be the same. However, it can be shown \cite{kittel,abrikosov} that Wiedemann-Franz law is valid for normal metals at low temperatures (approximately, at $T\leq 5$K see Ref.\onlinecite{Lorenzratios}), where all scattering processes are governed by impurities and defects, and at high temperatures $T\geq \Theta$ (here $\Theta$ is a characteristic Debye temperature), where high-energy phonons contribute the most to the electronic scattering. Experiment, indeed, shows (see, e.g., Ref.\onlinecite{kittel}) that at room temperature  pure metals demonstrate values of Lorenz number from $2.3\times  10^{-8} \frac{\mathrm{W}\cdot\mathrm{Ohm}}{\mathrm{K}^2}$ to $3.0 \times 10^{-8} \frac{\mathrm{W}\cdot\mathrm{Ohm}}{\mathrm{K}^2}$, i.e. within 20\% from the value predicted by Eqn.(\ref{eqn:Lorenz}).

While investigations of the Wiedemann-Franz law in pure metals are by now a well established classical matter (see Ref.\onlinecite{Lorenzratios} for a comprehensive review), Wiedemann-Franz law remains an active part of the modern physical research. Validity of the Wiedemann-Franz law at low temperatures is, in fact, a fundamental property of electronic fermi-liquid in a metal. Deviations from the Wiedemann-Franz law at low temperatures in the exotic conducting systems such as heavy fermion compounds \cite{hf1,hf2,hf3}, graphene, thin films or nanowires \cite{ultrapure,nanowire,1d,films} are carefully sought for and, if found, are considered as  signatures of new physical phenomena.

We present here a simple experimental setup to measure Lorenz number for a metal sample at room temperature. It can be partially assembled by the students during laboratory practice. Besides of providing an opportunity to measure combination of fundamental physical constants it also provides an opportunity to demonstrate several standard  ``tricks'' of physical experiment: 4-point resistivity measurement of small resistances, measuring different quantities on the same sample, compensation of the parasitic heat losses by setting sample environment to the same temperature. Experimental setup was tested for several metals, the order of magnitude for the Lorenz number is reproduced very reliably, its values for the studied metals are reasonably close to the theoretical value and to values derived from reference  values for resistivity and thermal conductivity.

\section{Experimental setup}
\begin{figure}
\includegraphics[width=\figurewidth]{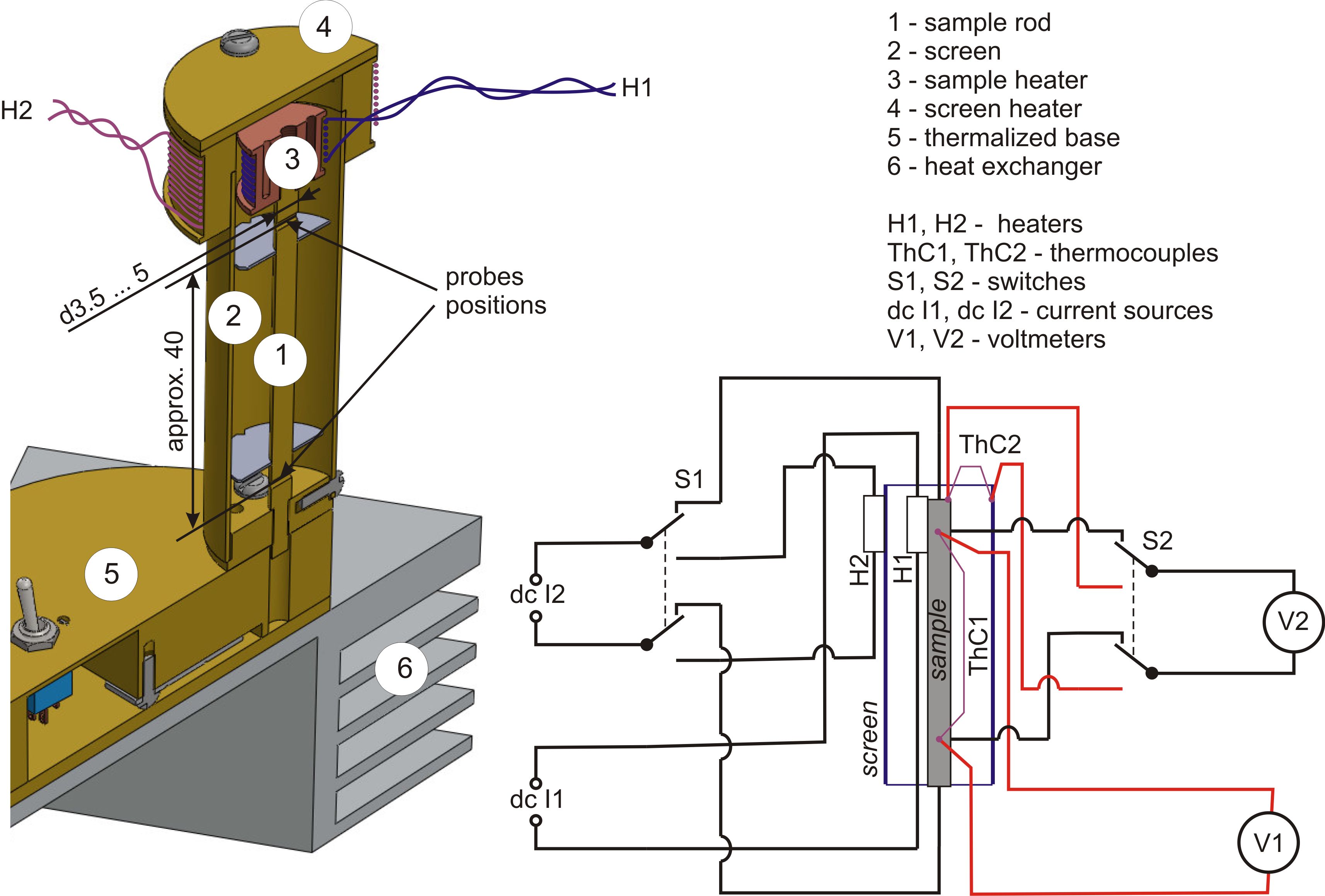}
\caption{ A sketch of the experimental cell and electric circuits of the experimental cell. Switches S1 and S2 on the circuitry scheme are in the ``resistivity measurement'' position. \label{fig:setup}}
\end{figure}
\begin{figure}
\includegraphics[width=\figurewidth]{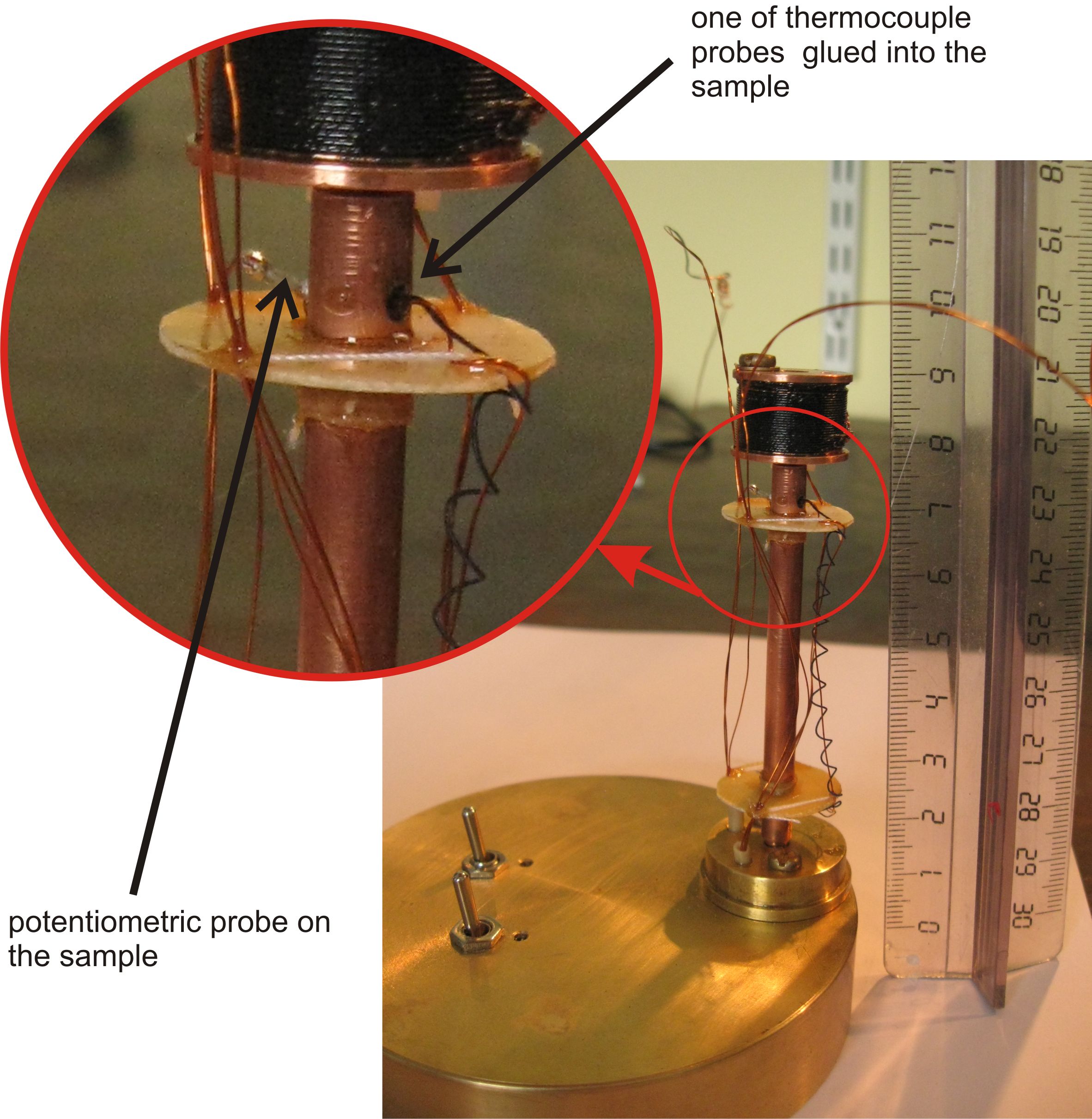}
\caption{ One of the experimental cells without compensating screen. Blown-up area shows one of the measurement points combining potentiometric contact and glued-in thermocouple end. \label{fig:setup-photo}}
\end{figure}

A sketch of the experimental cell and a photo of one of the cells are shown in Figs.\ref{fig:setup},\ref{fig:setup-photo}.

Lorenz number is defined as the ratio of  thermal and electric conductivities. Calculation of these  characteristics requires accurate determination of the sample geometry which introduces additional uncertainties to the determination of the desired quantity. However, if resistance and temperature difference are measured between the same points of the cylindric sample, then the sample geometry cancels out in the Lorenz number. Indeed, the resistance of such a sample is $R=l/(\sigma S)$ and the heat power transmitted through the sample crossection is $P/S=\kappa \Delta T/l$, here  $l$ is the sample length and $S$ is the sample crossection area. Then

\begin{equation}\label{eqn:L-integral}
L=\frac{\kappa}{\sigma T}=\frac{P}{\Delta T} \frac{R}{T}
\end{equation}

\noindent This makes one of the mentioned experimental ``tricks'': to set experiment so that sample geometry cancels.

The measurement of the resistivity is straightforward, we used a standard 4-point measurement scheme. We passed known DC electric current through the sample and  measured the voltage drop between potentiometric probes. This allowed to easily measure resistances as low as 50 microohms, as for copper sample in our setup. Measuring such a low resistance one has to take care about correctly zeroing the microvolmeter and to avoid issues with parasitic thermoelectric effects. These issues can be controlled by reversing current direction and fitting all data for voltage-current diagram as $U=R\times I+U_0$, $U_0$ being the parasitic voltage. We have found that in our experiment correction for this parasitic voltage (so present, see Fig.\ref{fig:test}) was negligible. It is also useful to remind  that the 4-point scheme allows to avoid various issues with the contact quality, which is of particular importance for the naturally oxide-covered aluminium, used as one of the samples.

To measure thermal conductivity we  applied controlled DC heating power to the upper end of the sample and thermalized its other end to the room temperature. It is of urgent importance then to minimize heat losses from the sample. To reach this goal without high vacuum or heavy thermal insulation we exploited two possibilities: first, we made crossection area of our sample large to increase ratio of the sample crossection to its sidewalls area; and, second, we surrounded our sample by a coaxial screen with the same temperature gradient. In this case, neglecting convective heat currents, there should be no heat exchange between the sample and the screen. To make the same gradient one end of the screen was thermalized to the same base as the cold (room temperature) end of the sample and other end was maintained at the same temperature as the hot end of the sample. Temperature difference along the sample was measured with copper-constantane thermocouple (sensibility $43\cdot 10^{-6}$ V/K at room temperature), its probes were attached at the same level as potentiometric probes for the resistivity measurements. Similar thermocouple was fixed between the hot ends of the sample and of the screen, its output was tuned to zero by setting appropriate heating power on the compensating screen. We have found it possible to balance temperatures of sample and screen hot ends within 0.25$^\circ$C quite easily.

Measuring thermal conductivity one has to wait sufficiently for thermal equilibrium. Characteristic time to reach thermal equilibrium of the cylindrical sample can be estimated as

\begin{equation}\label{eqn:thermalization}
    t_{\mathrm{eq}}\sim \frac{C^{(V)} l^2} {\kappa}
\end{equation}
\noindent here $l$ is the sample length, $C^{(V)}$ is the specific heat per unit volume, $\kappa$ is a thermal conductivity. At room temperature specific heat can be approximated by Dulong-Petit law $C^{(V)}=3R {\rho}/ {\mu}$, here $\rho$ is the mass density and $\mu$ is the molar mass. This yields for the 50 mm long copper sample, as used in our setup,   $ t_{\mathrm{eq}}\sim 20$ seconds. I.e., one has to wait 1-2 minutes to reach thermal equilibrium once the heating power is changed. As balancing of the sample and screen hot ends is done by trial and error, one need about 10-20 minutes in total to fine-tune heating power on the compensating screen.

Detailed sketch of the experimental setup is shown in Fig.\ref{fig:setup}. Sample is a 5 mm diameter cylinder rod made from copper, brass or aluminium alloy (duraluminium\footnote{Duraluminium contains approx. 4.5 mass\% of copper, approx. 1.5\% of Mg and some other components}) or a fragment of the 3.5 mm diameter aluminium wire. Distance between the probes was about 40-50 mm. Probes positions were formed by drilling a small diameter (1.6 mm) holes in the sample. For the copper and brass sample potentiometric probes were soldered into these holes and insulated thermocouple probes were glued in afterwards (see Fig.\ref{fig:setup-photo}), for the aluminium and aluminium alloy samples potentiometric electric contacts were formed by small diameter brass screws fixed in these holes, thermocouple probes were glued on the sample sidewalls on the same level. Compensating screen was manufactured from 24 mm diameter 1 mm thick brass tube. Thermalization of the device base was reached with commercial heat exchanger (approx. $150\times 70\times 40~\textrm{mm}^3$) without additional airflow.

To reduce amount of electronic equipment we switch electric circuits between resistance and thermal conductivity measurements. This allows to use two tunable current sources (2 A current limit) and two voltmeters (required sensitivity 1 mkV) only. We used GoodWill GDP-1831D power supplies, which have a built-in  current and voltage indicators, and Keithley 2000 multimeters. Scheme of the electric circuitry is shown at the Fig.\ref{fig:setup}. One of the power supplies (marked as ``dc I1'' in Fig.\ref{fig:setup}) and one of the voltmeters (V1) are always connected to the sample heater and to the sample thermocouple correspondingly. Second power supply (dc I2) and second voltmeter (V2) are switched between screen heater and screen-sample thermocouple or sample current and sample voltage drop correspondingly.
Sample and screen heaters with resistivity about 12 Ohms each were wound from the high-resistance 0.20 mm diameter manganine wire. Heating power during the experiment was about 5 W, commercial resistances with suitable characteristics can be used as well.

\section{Test results and discussion}
\begin{figure}
\includegraphics[width=\figurewidth]{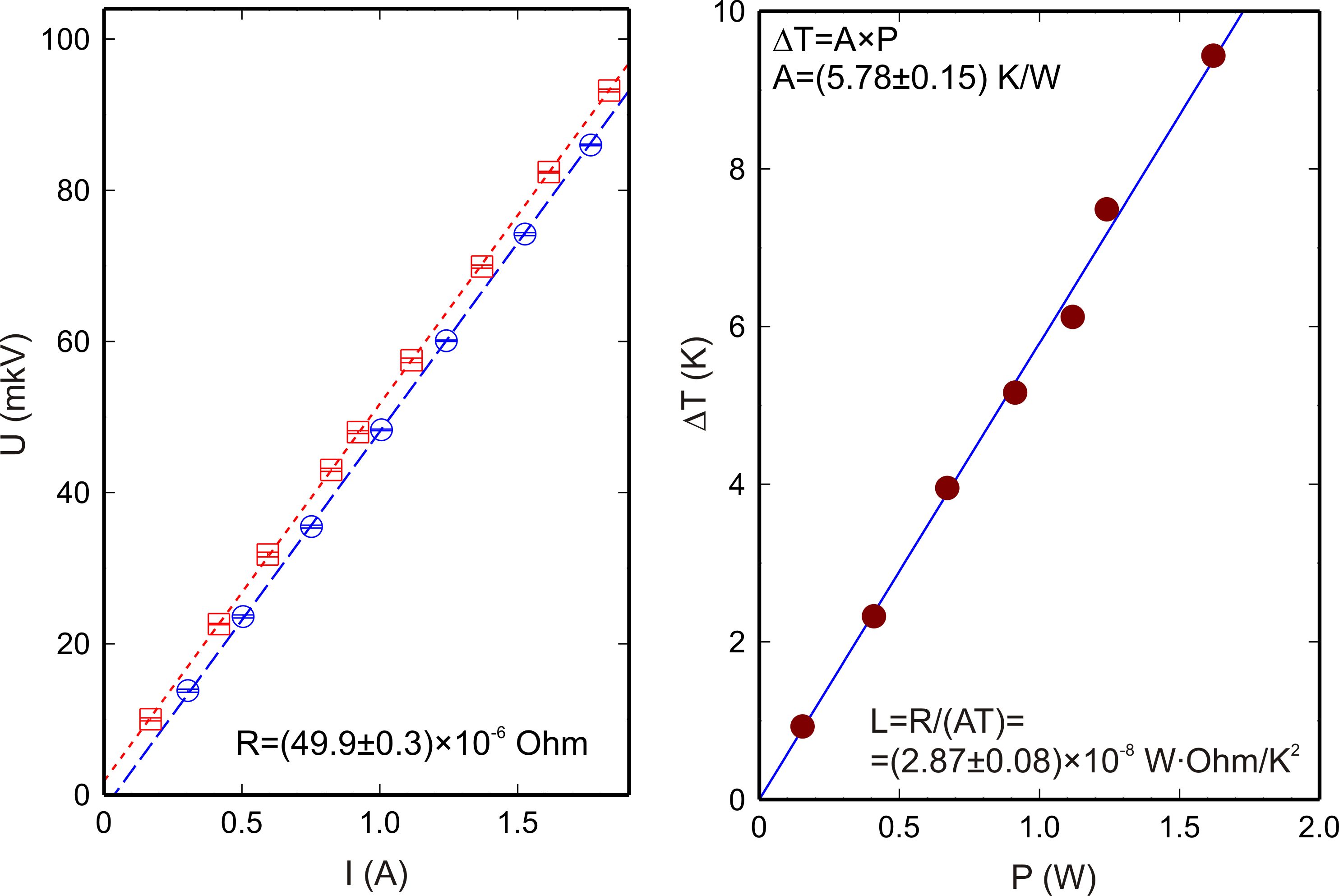}
\caption{(left panel) Voltage-current plot for the copper sample. Squares and circles --- experimental data for different polarities of the applied current, dashed lines --- fit by $U=R\times I\pm U_0$ law with $R=49.9\cdot 10^{-6}$ Ohm and $U_0=1.8$ mkV. (right panel) Temperature difference vs. heating power plot for the copper sample \label{fig:test}. Symbols --- experimental data, solid line --- linear fit.}
\end{figure}

We have tested our setup with two pure metals: copper and aluminium, and two alloys: brass and duralumin. Copper was studied in several cells differing by slight details of assembly.

Experimental procedure was as follows. First, we set switches into the resistivity measurement position and measured $U(I)$ characteristics of the sample up to 2 A current with two polarities of the DC current. By fitting $U(I)$ plot as $U=R\times I+U_0$ we determined total resistance of the sample. Second, we set switches to the thermal conductivity measurement position and heated sample hot end by applying DC heating power (about 0.5...1.0W for the first trial). As the temperature of the sample hot end rose we added DC heating power to the compensating screen to equalize temperatures of sample and screen hot ends (to zero thermocouple response). By fine-tuning the screen heating power it was possible to stabilize screen hot end within 0.1...0.2$^\circ$C from the sample hot end (thermocouple response less then 10 mkV). By repeating this procedure we measured dependence of the temperature drop along the sample vs. the sample heating power $ \Delta T(P)$ which was fitted afterwards as $\Delta T=A\times P$. Maximal temperature drop during the experiment was 10-15$^\circ$C, at maximum heating power screen hot end was warm on touch (about 50$^\circ$C) and thermalized base was slightly above the room temperature. Example of the measured characteristics for one of the copper samples is shown in Fig.\ref{fig:test}.

Lorenz number can be calculated from the measured quantities as $L=R/(A T)$ (compare with Eqn.(\ref{eqn:L-integral})). We take average temperature of 300K for this calculation as systematic errors due to the parasitic heat losses turns out to be far more important than the accuracy of the temperature determination. Determined Lorenz number values are shown in the Table \ref{tab:lorenz}. Measured values tends to overestimate Lorenz number, but order of magnitude is determined very reliably and reproductably.

To check main source of this overestimation we calculated from our data resistivity and thermal conductivity of the samples and compared it with the reference values (Table \ref{tab:rho-and-kappa}). Accuracy of probes positioning can lead to up to 5\% error in calculation of resistivity and thermal conductivity. Electrical resistivity of all samples was within 10\% of the reference values which can be at least partially ascribed to the uncertainties of the sample geometry and metal purity. Note, that resistances as low as $50 \cdot 10^{-6}$ Ohm are measured routinely during the experiment.

On the other hand, thermal conductivity was systematically overestimated with respect to  the reference values by 10--30\%. Part of this effect was a systematic heating power overestimation: when determining the heating power, we used power supply built-in voltmeter to measure voltage drop on the power supply contacts. This adds connecting wire resistance (about 0.5 Ohm) to the heater resistance (about 12 Ohm), ending up in the 4\% overestimation of thermal conductivity. This error can be avoided by using a separate voltmeter directly connected to the sample heater, but we consider usage of the power supply built-in voltmeter to be a reasonable trade-off in a teaching experiment. Another error source  are the parasitic heat losses being not  excluded completely in this simple setup. However, without compensating screen temperature drop along the sample decreases by a factor of two at the same heating power --- i.e. temperature gradient over the screen do compensate large part of the heat losses. This issue can be improved by increasing sample crossection, which reduces sidewalls to crossection area ratio, or by adding a more throughout thermal insulation, in particular between the sample heater and the compensating screen. This is, again, a matter of a trade-off between the experiment simplicity and the desired accuracy. 

Examples of other experimental setups for similar experiments were reported recently. \cite{example1,example2} A strong point of our experiment is that both electric and thermal properties were measured on exactly the same sample (as it is also done in a ``serious science'' experiment of Ref.\onlinecite{films}, where Wiedemann-Franz law was tested in a 100 nm thick films) and that an absolute value of the Lorenz number, which is a combination of fundamental constants, is derived.

\begin{table}
\caption{Measured values of Lorenz number for different materials and their reference values. Values from  Ref.\onlinecite{kittel} correspond to 0$^\circ$C, values from Ref.\onlinecite{physquant} are calculated from the reference values for specific resistivity and thermal conductivity at 0$^\circ$C or at 20$^\circ$C, values from Ref.\onlinecite{emat} correspond to 20$^\circ$C, data from Ref.\onlinecite{Al-alloy} are extrapolated to 0$^\circ$C. \label{tab:lorenz}}
\begin{tabular}{c|c|c}
Sample& measured $L$ $\left(10^{-8} \frac{\mathrm{W}\cdot\mathrm{Ohm}}{\mathrm{K}^2}\right)$& reference $L$ $\left(10^{-8} \frac{\mathrm{W}\cdot\mathrm{Ohm}}{\mathrm{K}^2}\right)$\\
\hline
Cu-1&$2.95$&\multirow{4}{*}{$\left.\begin{array}{c} \\ \\ \\ \\\end{array}\right\}$2.23 \cite{kittel}; 2.28\cite{physquant}, 2.30\cite{emat}}\\
Cu-2&$2.45$&\\
Cu-3&$2.70\pm 0.06$&\\
Cu-4&$2.87\pm 0.08$&\\
yellow brass&$3.4\pm0.2$&2.50 \cite{emat}\\
Al&$3.05\pm0.08$&2.17 \cite{physquant}, 2.18 \cite{emat}\\
Al alloy&$3.02\pm0.1$& 2.3 \cite{Al-alloy}\\
\end{tabular}
\end{table}

\begin{table}
\caption{Resistivities and thermal conductivities of the studied metal samples and their reference values \label{tab:rho-and-kappa}}
\begin{tabular}{c|c|c|c|c|c|c|c|c}
Sample&$d$&$L$&$R$&$A=\Delta T/P$& \multicolumn{2}{|c|}{$\rho$ $\left(10^{-8}\mathrm{Ohm}\cdot\mathrm{m}\right)$}&\multicolumn{2}{|c}{$\kappa$ $\left(\frac{\mathrm{W}}{\mathrm{m}\cdot\mathrm{K}}\right)$}\\
&(mm)&(mm)&($10^{-6}$ Ohm)&(K/W)&measured&reference&measured&reference\\
\hline
Cu-1&5&&60&7.5&2.2&\multirow{4}{*}{$\left.\begin{array}{c} \\ \\ \\ \\\end{array}\right\}$1.70\cite{kittel,emat}, 1.55\cite{physquant} }&407&\multirow{4}{*}{$\left.\begin{array}{c} \\ \\ \\ \\\end{array}\right\}$401\cite{physquant}}\\
Cu-2&5&52&49.5&6.11&1.87&&433&\\
Cu-3&5&50&47.7&5.80&1.9&&440&\\
Cu-4&5&50&49.4&5.78&1.95&&440&\\
yellow brass&5&50&155&15.1&6.1&6.4\cite{metall}&170&110\cite{physquant}\\
Al&3.5&42&135&14.7&3.1&2.74\cite{kittel}, 2.50\cite{physquant}, 2.7\cite{emat}&300&237\cite{physquant}\\
Al alloy&5&50&148&15.3&5.8&5\cite{Al-alloy}&170&130\cite{physquant,Al-alloy}\\
\end{tabular}
\end{table}

\section{Conclusions}

We demonstrated a simple experimental setup to check validity of Wiedemann-Franz law in the teaching experiment. It was tested with several common metals and alloys. We have found that order of magnitude of the Lorenz number was reliably reproduced by our setup. In the same time obtaining an accurate value of the Lorenz number remains a challenge to the student patience, accuracy and experimentalist skills.


\begin{thebibliography}{99}

\bibitem{kittel} C.Kittel, Introduction to Solid State Physics, John Wiley and Sons (2005)
\bibitem{wf} R.Franz and G.Wiedemann, Annalen der Physik (in German), \textbf{165}, 497 (1853)
\bibitem{lorenz} L.Lorenz, Annalen der Physik (in German), \textbf{223}, 429 (1872)
\bibitem{abrikosov} A.A.Abrikosov, Fundamentals of the Theory of Metals, Elsevier Science Pub. Co. (1988)
\bibitem{Lorenzratios} J.G.Hust, L.L.Sparks, Lorenz Ratios of Technically Important Metals and Alloys, NBS Technical Note No.634 (1973)
\bibitem{hf1} R. Mahajan, M. Barkeshli, and S. A. Hartnoll, Physical Review B \textbf{88}, 125107 (2013)
\bibitem{hf2} J. Paglione, M. A. Tanatar, D. G. Hawthorn, F. Ronning, R.W. Hill,
M. Sutherland, L. Taillefer, and C. Petrovic, Physical Review Letters \textbf{97}, 106606 (2006)
\bibitem{hf3}J.-Ph. Reid, M. A. Tanatar, R. Daou,  Rongwei Hu,  C. Petrovic,  and Louis Taillefer, Physical Review B \textbf{89}, 045130 (2014)
\bibitem{ultrapure} A. Principi and G. Vignale, Physical Review Letters \textbf{115}, 056603 (2015)
\bibitem{nanowire} F. V\"{o}lklein, H. Reith, T. W. Cornelius, M. Rauber and
R. Neuman, Nanotechnology \textbf{20},  325706 (2009)
\bibitem{1d} N. Wakeham, A.y F. Bangura,  Xiaofeng Xu, J.-F. Mercure, M. Greenblatt and N. E. Hussey, Nature Communications \textbf{2}, 396 (2011)
\bibitem{films} A. D. Avery, S. J. Mason, D. Bassett, D. Wesenberg and B. L. Zink
Physical Review B \textbf{92}, 214410 (2015)
\bibitem{physquant}  (editors) I.Grigoriev, E.Z.Meilikhov, Handbook of physical quantities, Boca Raton : CRC Press, (1997)
\bibitem{emat} L.A.A.Warnes, Electronic Materials, Macmillan Education Ltd. (1990)
\bibitem{metall} M.Bauccio (ed.), ASM Metals Reference Book, ASM International (1993)
\bibitem{Al-alloy} R.A.Overfelt, S.I.Bakhtiyarov, R.E.Taylor, High Temperatures-High Pressures, \textbf{34}, 401 (2002)
\bibitem{example1} A. Clark, L.N. Jacobson and P. Polley, Journal of Physical Science and Application \textbf{3}, 328  (2013)
\bibitem{example2} K. Meehan, R. Hadfield and A. Phillips,  2015 ASEE Annual Conference and Exposition, Seattle (2015) (doi:10.18260/p.23450)
\end{thebibliography}
\end{document}